\documentclass[slac_one]{revtex4}
\usepackage{graphicx}
\usepackage{fancyhdr}
\begin{document}

\title{A Global Event Description using Particle Flow with the CMS Detector } 

\author{J.~Weng}
\affiliation{Institute for Particle Physics, ETH Zuerich, Switzerland}

\begin{abstract} The CMS Detector consists of a large volume silicon tracker immersed in a
high four Tesla magnetic field, together with a high resolution/granularity
electromagnetic calorimeter and a nearly full solid angle coverage hadronic
calorimeter. Particle flow reconstruction provides a global pp-collision
event description by exploiting the combined information across all CMS
sub-detectors, optimizing the reconstruction and the identification of each
particle (photons, electrons, muons, unstable neutral hadrons, charged
hadrons and neutral hadrons) in an event. This summary introduces the CMS
particle flow algorithm, discusses the challenges associated with the LHC
environment, and presents some first example results in the context of hadronic decays of taus as
well as missing transverse energy.
\end{abstract}

\maketitle

\thispagestyle{fancy}

\section{Particle Flow Overview} 
The particle-flow reconstruction algorithm aims at providing a global ({\it i.e.}, complete and unique) event description 
at the level of individually reconstructed particles, with an optimal combination of the information coming from all CMS
subdetectors. The reconstructed and identified individual particle list includes muons, 
electrons (with individual reconstruction and identification of all Bremsstrahlung photons), photons (either unconverted or
converted), charged hadrons (without or with a nuclear interaction in the tracker material), as well as stable and unstable 
neutral hadrons. These particles can be non isolated, and even originate from a intricate overlap of reconstructed charged particles, 
ECAL and HCAL energy clusters, and signals in the muon chambers. 

The particle-flow algorithm starts with reconstruction performed independently within each CMS sub-detector: clustering is 
conducted separately in each of the electromagnetic (ECAL) and hadronic (HCAL) calorimeters and track reconstruction takes 
place in the combined silicon and pixel tracker system.  

The efficiency of tau identification is directly proportional to the tracking efficiency. In addition, the energy of 
each charged particle missed by the tracking algorithm is determined with the limited hadron calorimeter resolution, and 
its direction is biassed by the large magnetic field. Conversely, fake tracks may rapidly dominate the jet energy and 
angular resolutions, and may dramatically increase the QCD background to tau jets. An iterative tracking algorithm was therefore 
developed in the context of particle flow to improve the overall CMS tracking efficiency and fake rate.The iterative 
tracking strategy is as follows. First, tracks with a very pure seeding algorithm (such as the requirement of three 
hits in the pixel detector, the requirement of a minimum number of hits in the silicon tracker, with tight vertex 
constraints) are reconstructed. This first iteration leads to a moderate efficiency and a very small fake rate. The 
next steps proceed by cleaning hits used in the previous iteration and by progressively loosening track quality 
criteria, thus increasing the tracking efficiency without noticeably changing the fake rate. With this technique, 
charged particles with as little as three hits and a transverse momentum as small as 300\,MeV/$c$ can be reconstructed 
with a larger efficiency and a fake rate similar to those. The original one-step reconstruction which required at least 
eight hits and a transverse momentum in excess of 0.9\,GeV/$c$

Tracks originating from secondary vertices (e.g., from the nuclear interaction of a hadron or a photon conversion in the tracker
material, or from the decay of unstable neutral hadrons) are seeded with specific algorithms from the hits remaining after the 
first-steps cleaning. 

The particle-flow calorimeter clustering algorithm is conceptually identical for ECAL crystals and HCAL cells. Clusters are seeded 
by local energy maxima in topologically-connected sets of cells. The energy of any nearby crystal/cell is then shared between 
individual clusters according to a crystal/cell-to-cluster distance, allowing for better spatial resolution of overlapping clusters 
and with less biased cluster position measurements than the ECAL clustering algorithms developed, e.g., for isolated electrons and 
photons. 
Links between ECAL clusters, HCAL clusters, and tracks are made, depending on the spatial and energy compatibility between 
the clusters and/or tracks.  The particle-flow algorithm proceeds in stages, associating clusters and tracks with a newly reconstructed 
particle at each progressive stage.  First, tracks and clusters identified as being associated with hits and segments in the muon 
chambers are tagged as muons and removed from the list of unassociated objects. Next, tracks and clusters identified as being associated 
with electrons (and all the possible individual Bremsstrahlung photons) are tagged and removed from further processing. 
Next, in the case of an HCAL cluster linked to a track, the calibrated HCAL cluster energy is compared with the track momentum.  
If the cluster energy is compatible with the track momentum, a charged hadron is created with energy determined from a weighted 
average of the track momentum and the cluster energy. If the difference between the cluster energy and the track momentum is 
significant (with respect to the expected calorimeter energy resolution and the measured track momentum uncertainty), a neutral 
hadron is created out of the excess cluster energy.  In the case where an ECAL cluster and an HCAL cluster are linked together 
with a track, the calibrated combined energy of the ECAL cluster and HCAL cluster is compared with the track momentum. If the 
combined ECAL and HCAL calorimeter energy is compatible with the track momentum, a charged hadron is created as explained above; 
otherwise either a neutral hadron or a photon is created out of the excess calorimeter energy, depending on a multivariate analysis 
of the track momentum, the relative energy deposits in ECAL and HCAL, the cluster-track link quality and the transverse cluster shapes.
After removing these tracks and clusters from the list of unassociated objects, only clusters not linked to any track remain uncleaned 
from the event. Any such ECAL clusters are assumed to be photons and any such HCAL clusters (or HCAL clusters linked with ECAL clusters) 
are assumed to be neutral hadrons.

\begin{figure}[htp]
      \includegraphics[angle=90, width=0.40\textwidth]{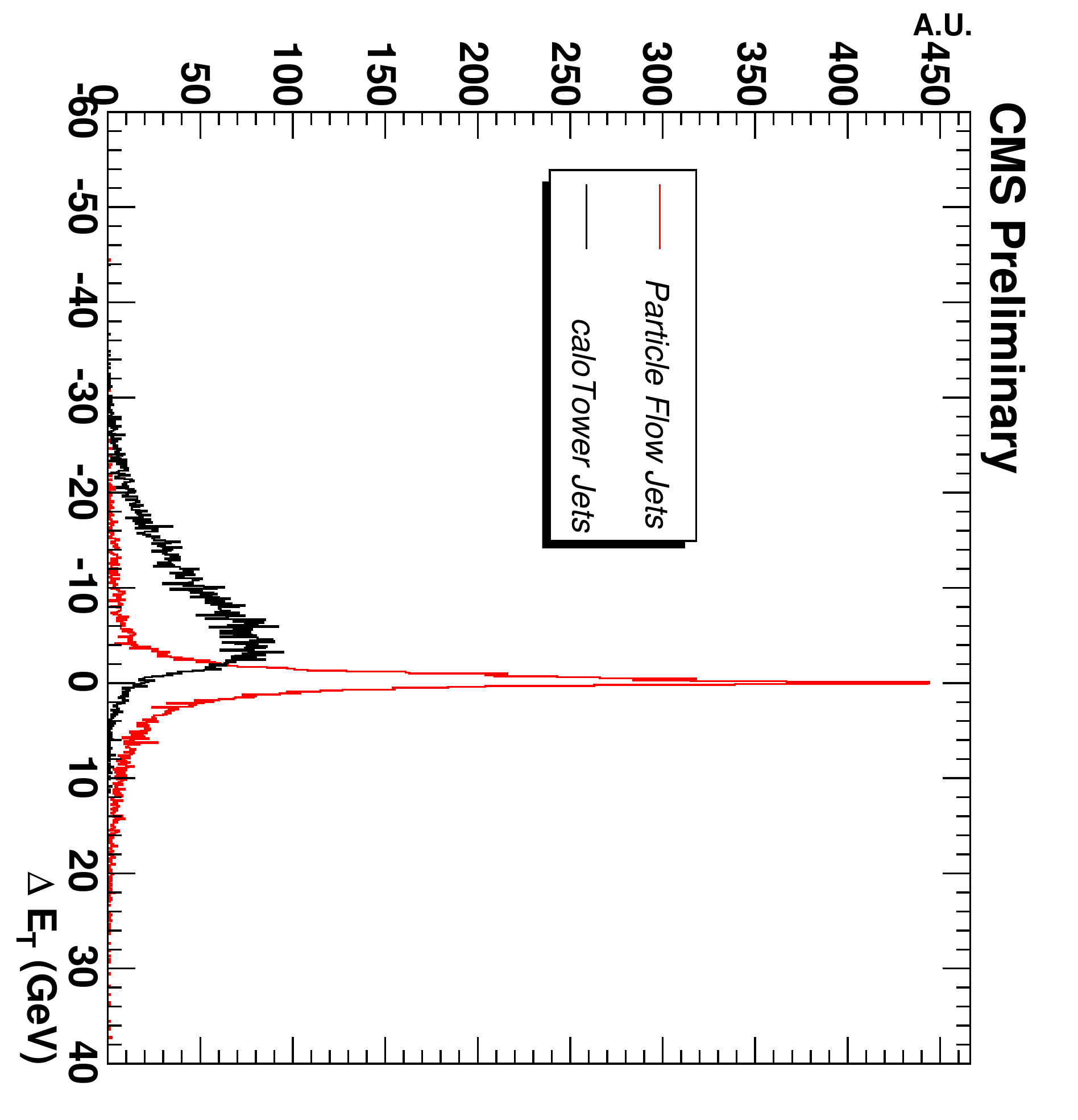}
      \includegraphics[angle=90, width=0.40\textwidth]{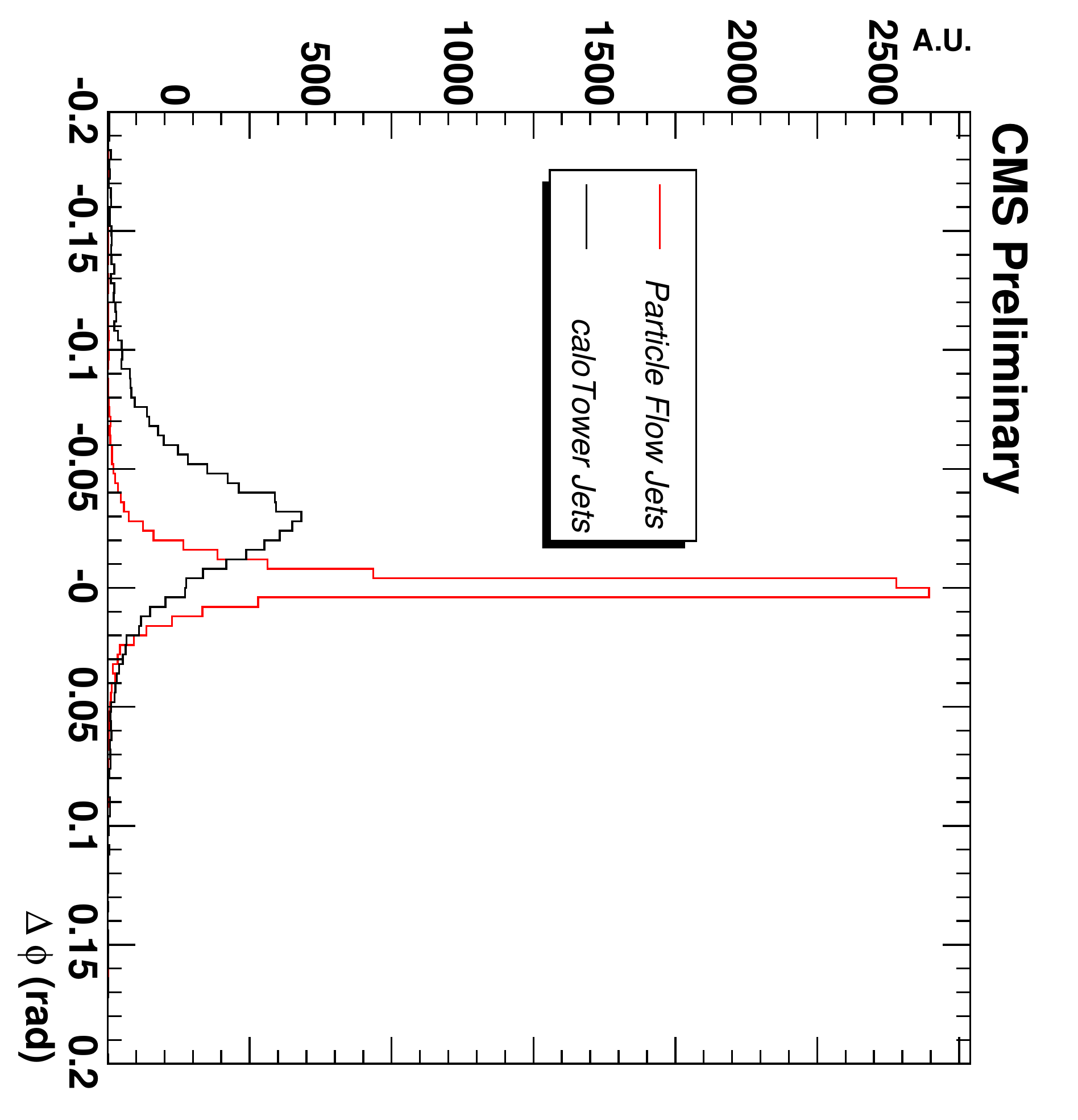}
   \caption{Comparison between particle-flow reconstruction (red) and calorimeter-based (black) reconstruction of single taus with 
$p_{\rm T} = 50$\,GeV/$c$ (typical of a Z decay). Left: Difference, in GeV, between the reconstructed and the true visible transverse
momentum; Right: Difference, in radian, between the reconstructed and the true azimuthal angle.
	Jet energy corrections are not applied in both cases.}
   \label{fig:pflow}
\end{figure}

The complete list of particles (muons, electrons, photons, charged hadrons, neutral hadrons) may then be used 
to derive composite physics objects, such as clustering into jets with standard jet algorithms.  Tau jets reconstructed with this approach benefit
from both the improved energy and angular resolution with respect to the calorimeter-based algorithm (Fig.~\ref{fig:pflow})  
and the depth of information available describing each individual particle in the jet.

The limited energy and angular resolution in Figure~\ref{fig:pflow} of the calorimeter-based jets is 
dominated by the hadron calorimeter resolution and granularity. As the tau decay products are mostly photons and charged pions, 
the particle-flow-based jets benefit fully from the tracker and electromagnetic calorimeter superior resolutions.
The azimuthal-angle bias of the calorimeter-based jets is caused by the large axial magnetic field. (Only $\tau^-$ are simulated). 
In the particle-flow-based jets, the directions of the charged hadrons are measured from their momenta determined at the 
primary vertex by the tracker, hence are not affected by the magnetic field. Finally, the bias in the calorimeter-based 
jet energy cannot be corrected by the regular jet-energy calibration which is derieved from QCD jets and would overshoot the tau true energy.
Instead, particle-flow-based jets are calibrated by construction, with the use of the accurate charged-particle-momentum and 
photon-energy determination. These refinements allow backgrounds otherwise misidentified as a hadronically 
decaying tau to be more effectively rejected. More details can be found in~\cite{ref:pf}.

\section{Tau corrections to Missing transverse energy}
\begin{figure}[htp]
      \includegraphics[ width=0.50\textwidth]{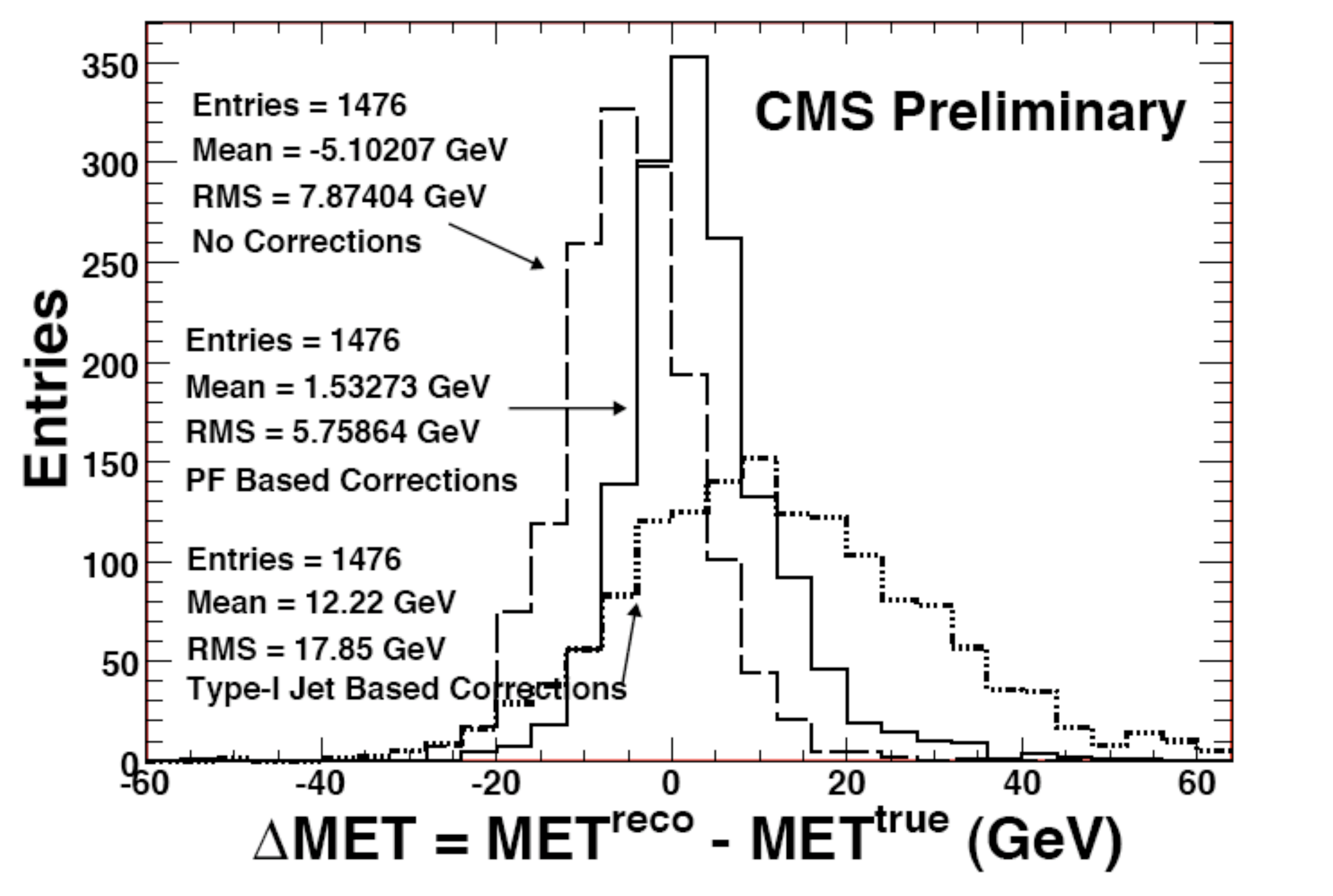}
  
   \caption{Distribution of reconstructed minus true missing transverse energy ($E_{T_{reco}} -E_{T_{true}}$) for hadronically decaying tau
 leptons in W (+0 jets) $\rightarrow \tau_{\rm hadronic} \nu$ events for three cases: 
no correction (dashed line), standard jet correction (dash-dotted), and Particle Flow-based (solid) correction.}
   \label{fig:pflowmet}
\end{figure}

Particle Flow based tau corrections has been also used to improve the reconstructed missing $E_T$ (MET).
Usually, jet energy scale corrections are applied here as correction. 
However, tau jets are substantially different from ordinary jets, making standard jet corrections inappropriate for taus. 
These differences arise from the fact that hadronic taus typically have a small number of fairly energetic 
particles while QCD jets of the same energy have higher multiplicity and a larger fraction of energy carried by soft particles. 
Applying standard jet corrections to hadronic tau jets results in a significant over-correction of the missing $E_T$. 
This is illustrated in Figure~\ref{fig:pflowmet} which shows the distribution of the reconstructed missing energy minus the 
generated missing energy ($E_{T_{reco}} -E_{T_{true}}$) for visible hadronic tau energy in W events in the cases without
 jet energy scale corrections (dashed line) and when standard jet corrections from Monte Carlo are applied (dash-dotted line).
Neither case yields a satisfactory description of the jet $p_T$.
We define the correction procedure for hadronic tau jets, which satisfy certain identification requirements.
The MET correction is then calculated as follows:
\begin{equation}
\Delta \vec MET = \sum \vec{ E_T}^{\rm cal \, jet 0.5}  - \vec E_T^{\rm PF \tau}
 \end{equation}

The first term heer can be approximated by the transverse energy of a jet obtained using standard conebased
jet clustering algorithm with the cone size of 0.5 (modulo possible difference in energy thresholds used
in calculating raw MET and in the jet energy summation). The second term should be approximated by the best
available measurement of the tau energy using particle flow (PF). Corrections from underlying event(UE) and pile-up (PU)
are estimated to be small for early running conditions and will be determined later.  
To gauge the performance of this correction and disentangle effects associated with other corrections (e.g. for
jets, electrons and muons), we select a sample of W events and require that there be no additional jets
with $E_T > 5$ GeV. While this is a harsh requirement, it is effective in removing other effects associated with
mismeasurements of recoil jet energies. Figure~\ref{fig:pflowmet} shows the MET distribution for three cases:
No correction (dashed line), standard jet correction (dash-dotted) and PF-based (solid) correction. The PF based
calculation is seen to yield the result with the smallest bias and the best resolution. The absence of a visible
bias confirms that effects of the UE are small. More details can be found in~\cite{ref:met}.


\begin{thebibliography}{9}   
\bibitem{ref:pf}
CMS collaboration, ``CMS Strategies for tau reconstruction and identification using particle-flow techniques", 
CMS PAS PFT-08-001.
\bibitem{ref:met}
CMS collaboration, ``Missing Et performance in CMS", 
CMS PAS JME-07-001. 


\end{thebibliography}
\end{document}